\documentclass{article}
\usepackage{graphicx}

\begin{document}  
                             
\noindent {\large Wavelet Correlation Coefficient of
'strongly correlated' financial time series}
\vskip1.5cm 
\noindent {\bf Ashok Razdan}

\noindent {\bf Nuclear Research Laboratory }

\noindent {\bf Bhaba Atomic Research Centre }

\noindent {\bf Trombay, Mumbai- 400085 }
\vskip 1.5cm
\noindent {\bf Abstract :}

In this paper, wavelet concepts are used to study two 
'strongly correlated' financial time series. Apart from obtaining
wavelet spectra,
We  also calculate wavelet correlation coefficient and 
show that strong correlation or strong anti-correlation 
depends on scale.

\vskip 0.5cm 
\noindent {\bf Introduction:}

In recent times lot of activity has been witnessed in the the field of
econophysics [ 1]. Various concepts and methods of Physics and Mathematics have
been applied to study financial time series both for long range and short
range studies. The probability distribution function for money
has been found to  follow
Boltzman Gibbs laws  displaying typical equilibrium state of maximum
entropy [2].
It has been found that financial time series (like stock
indices, currency exchange rates etc) have either fractal or multi-fractal
features [3,4 and references therein].  Earlier it was assumed that stock market returns follow random
walk model and indices are normally distributed [5]. However, it been 
proved in recent times that indices are not normally 
distributed but have higher peaks and
fatter tails [6,7,8]. Such distributions are known as stable paretians which have
infinite or undefined variance. The presence of fatter tails indicate
memory effects which arise due to non-linear stochastic processes. Another
important feature that has been observed in recent times is that stock
market indices can be represented by fractional Brownian motion instead of
classical Brownian motion [3]. The similarity between fluid turbulence
and financial markets is well known[ 9 ,10 ].  The information transfer in
financial markets is similar to energy flow in hydrodynamics [ 11].

\noindent {\bf Correlation studies in Indian stock markets:}

The cross-correlation coefficient r which
is a measure of linear association between two variables is defined as
\begin{equation}
r=\frac{\sum_{i=1}^{n}(X_i - \overline{X})(Y_i -\overline{Y})}
    {\sqrt{\sum_{i=1}^{n}(X_i -\overline{X})^2)
    \sum_{i=1}^{n}(Y_i -\overline{Y})^2)}}
\end{equation}
A positive value of coefficient r indicates, that as one value increases
the other tends to increase whereas a negative value indicates
as one variable increases the other tends to decrease.

\begin{figure}
\begin{center}
\caption{BSE and NSE indices for the year 2000} 
\includegraphics[angle=270,width=6.0cm]{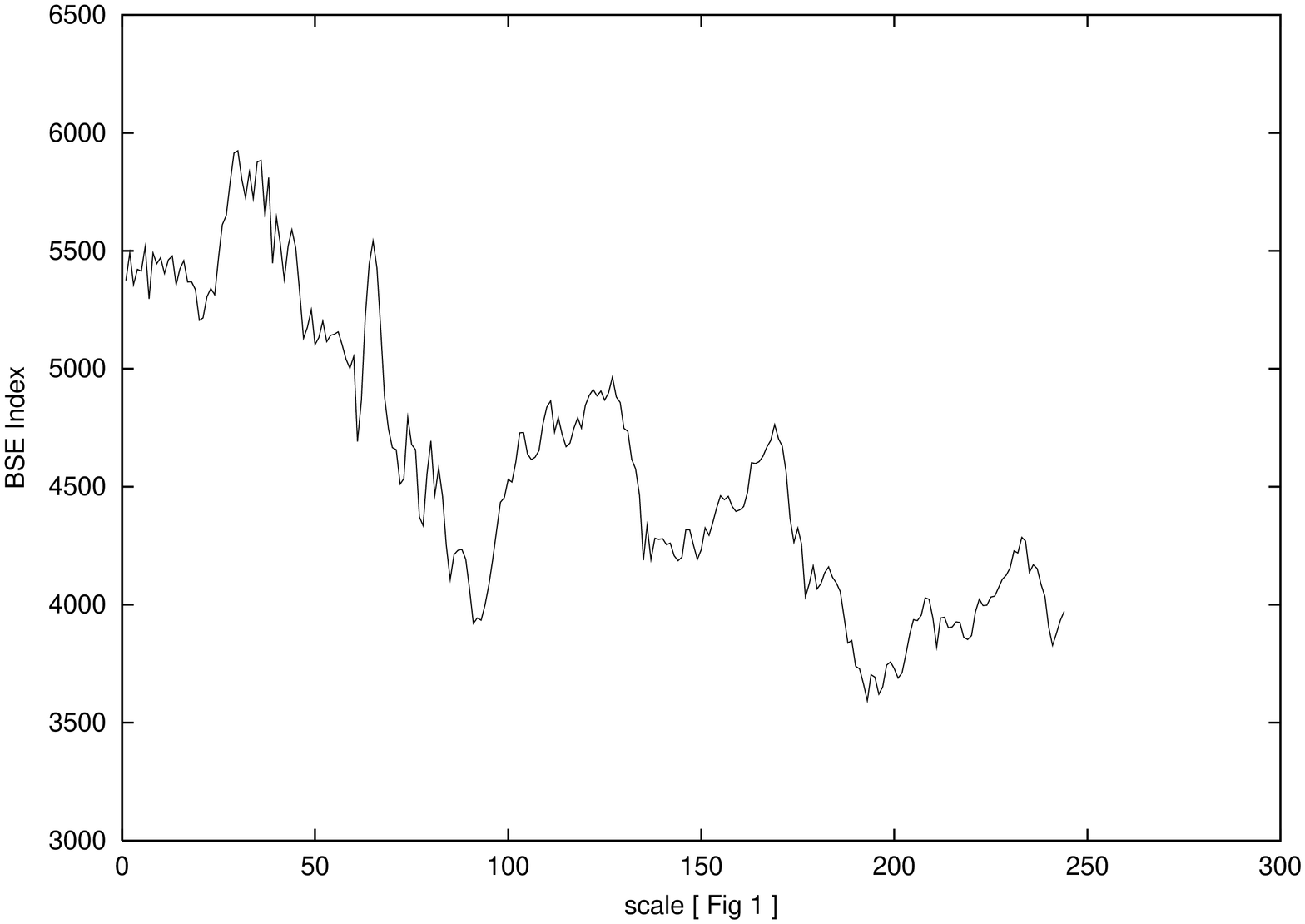} \includegraphics[angle=270, width=6.0cm]{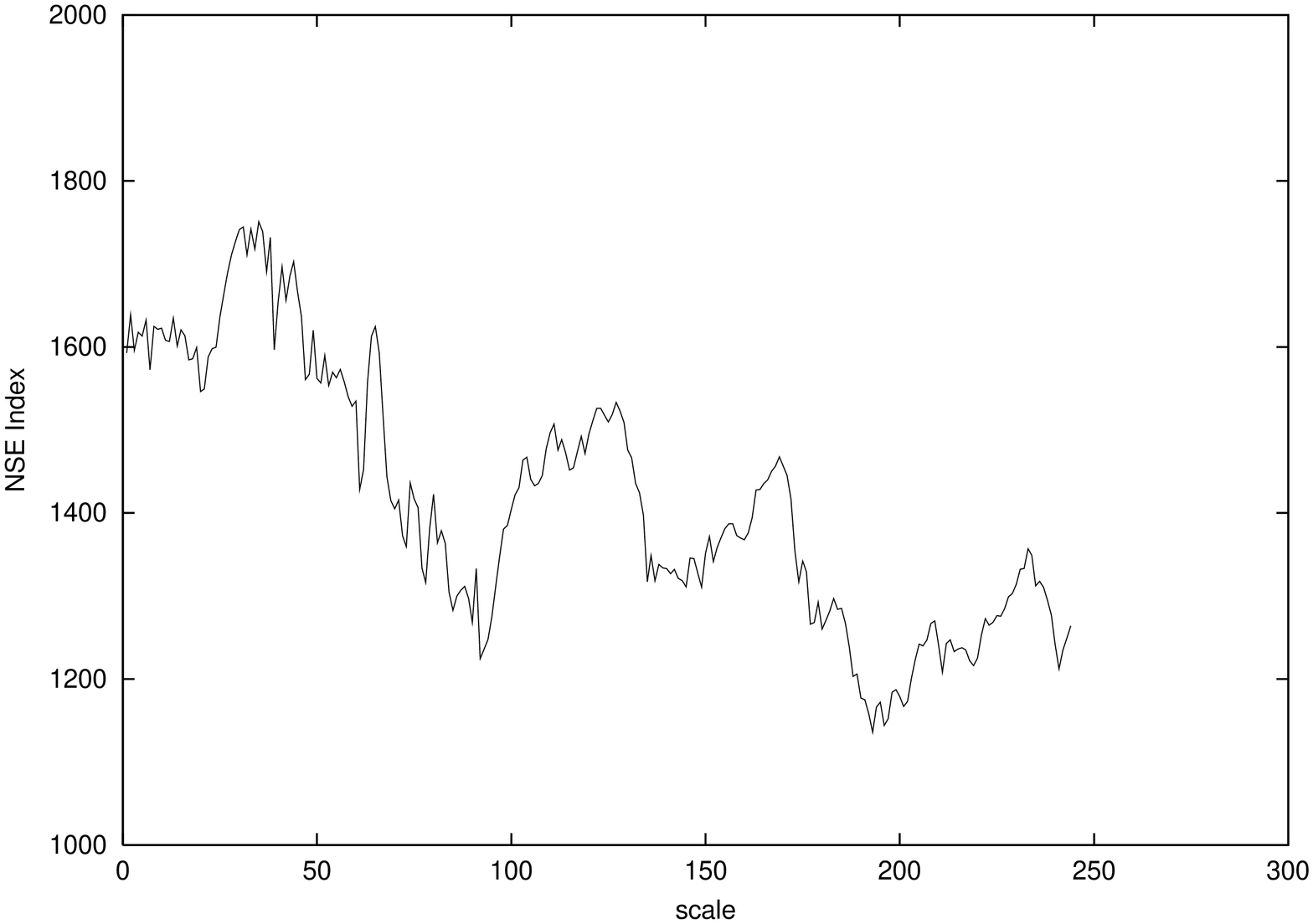}
\end{center}
\end{figure}

In this paper we study correlation  between
Bombay stock exchange Index ( BSE) and National stock  exchange index (NSE).
Both of these stock exchanges belong to India and open and close at the
same time i.e. they are synchronous. BSE index and NSE index are known to
be strongly correlated.
Bombay stock exchange is the oldest stock exchange in whole of Asia. It was
started in the year 1875. BSE index is market capitalization weighted index
of 30 stocks of sound Indian and multinational  companies. National stock exchange was established
to provide access to investors from all across India. NSE started equity
operations in November 1994 and operations in derivatives in June 2000.
NSE index also known as NIFTY is determined from 50 stocks of companies
taken from 23 sectors of economy.
Figure 1 shows BSE  index and NSE index for the year 2000.  It is clear from
figure 1 that BSE and NSE indices are very similar and correlated.
The calculated value of correlation coefficient  ( equation 1) between BSE
and NSE indices is r =0.993100822. This value of r also indicates that BSE
index is strongly correlated with NSE index. This is also evident from figure 2
which shows BSE and NSE indices highly correlated.  

\begin{figure}
\begin{center}
\includegraphics[angle=270,width=7.0cm]{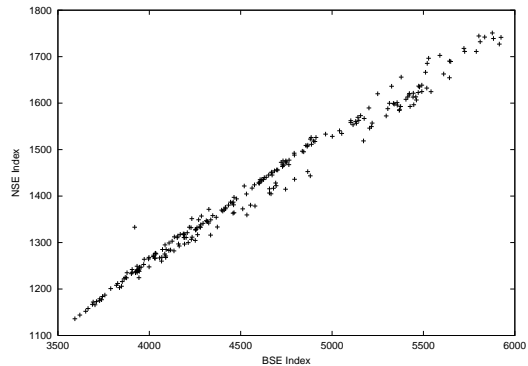}
\caption{Scatter of BSE  and BSE indices} 
\end{center}
\end{figure}
\vskip 1cm

In recent times wavelets have been used to quantify various parameters
in various scales. 
In this paper we use wavelet based correlation coefficient to study correlation between
Bombay stock exchange Index ( BSE) and National stock  exchange index (NSE).

\noindent{ \bf Wavelet correlation  coefficient:}

Wavelet analysis is being used increasingly to study given structures
in different scales [12,13,14].  Wavelets can detect both the location and a
scale of a structure. Wavelets are parameterized both by scale a$>$ 0
(dilation parameter) and a translation parameter b $(-\infty < b < \infty)$
such that
\begin{equation}
\Psi_a,b=\frac{\psi(x-b)}{a}
\end{equation}

The wavelet domain of one dimensional function $\Psi$ is rather two dimensional
in nature; one dimension corresponds to scale and other to translation.
The continuous wavelet transform for one dimensions is defined as
\begin{equation}
w(a,b)= \int dx f(x) \frac {\psi^{*}(|x-b|)}{a}
\end{equation}
where a is the scale. Here f(x) is one dimensional function and
$\psi^{*}$ (* is complex conjugate) is the analyzing
wavelet or also known as mother wavelet.
We choose a Mexican hat wavelet as a  possible analyzing wavelet [20]
\begin{equation}
\psi\frac{(|x-b|)}{a}=\frac{1}{(2\pi)^{0.5} a} (2- \frac{|x-b|^2}{a^2}) exp(- \frac{|x-b|^2}{2a^2})
\end{equation}
Mexican wavelet is an isotropic wavelet having minimum number of oscillations.
An important property of wavelet $\psi$ to be used as analyzing wave,it must
have zero mean value i.e. $ \int \psi(x)$ dx=0. $\psi$ is also required
to be orthognal to some lower order polynomials i.e.
\begin{equation}
\int_{-\infty}^{\infty}  x^{m} \psi(x) dx =0
\end{equation}
where  $0 \le m \le n $ . Here n is the upper limit related to the order
of wavelet.

\begin{figure}
\begin{center}
\includegraphics[angle=270,width=7.0cm]{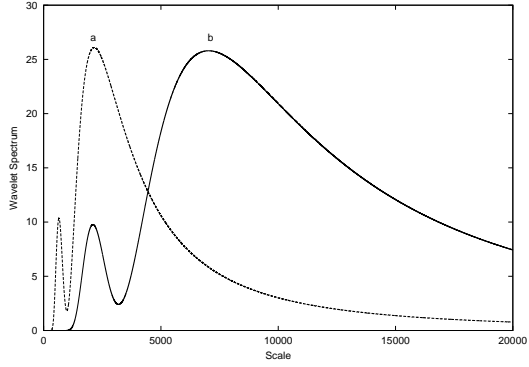}
\caption{Wavelet Spectrum of NSE ( curve a) and BSE (curve b) indices} 
\end{center}
\end{figure}

A wavelet transform is the inner-product of the function with scaled and
translated wavelet. A wavelet transform decomposes a given function into
coefficients from which original function can be reconstructed.
By using equation (3) wavelet spectrum can be defined as [15]
\begin{equation}
M(a)=\frac{1}{a} \int_{\infty}^{\infty} |w(a,b)|^{2} dx
\end{equation}
Wavelet spectrum has a power law behavior
\begin{equation}
M(a) \sim  a^{\lambda}
\end{equation}
Wavelet spectrum M(a) defines energy of wavelet coefficients  for scale ' a'.
Here $\lambda$ is an exponent, value of which is decided by power of a.
Wavelet cross-correlation coefficient has been defined as [25]
\begin{equation}
r_w (a)=\frac{ \int w_1(a,b) w_2^{*}(a,b) db}{(M_1(a) M_2(a))^{\frac{1}{2}}}
\end{equation}
The relationship between correlation coefficient and wavelet correlation
coefficient can be written as
\begin{equation}
r=\frac {\int r_w(a) ( M_1(a) M_2(a) )^{\frac{1}{2}} a^{-1} da}{\int (M_1(a) a^{-1} da \int M_2(a) a^{-1} da)^{\frac{1}{2}}} 
\end{equation}
The concept of wavelet correlation coefficient has been applied to study scale
dependence in various galaxies [15].

\begin{figure}
\begin{center}
\includegraphics[angle=270,width=7.0cm]{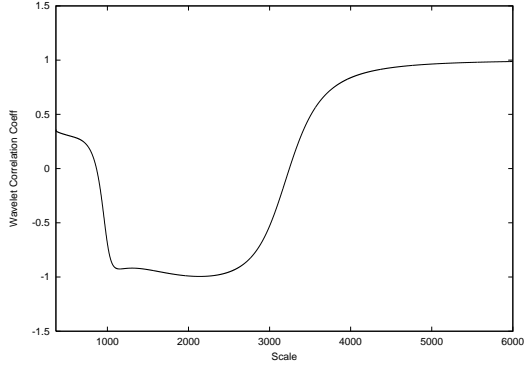}
\caption{Wavelet Correlation Coefficient between BSE and NSE indices} 
\end{center}
\end{figure}

\noindent {\bf  Results and Discussion:}

In figure 3 we have plotted
wavelet spectra of BSE (figure a) and NSE (figure b) indices. The wavelet spectra
of two indices have similar features.
Both spectra have two maxima and one minima. Maxima of both
spectra corresponds to maximum values of indices and minimum in the
spectra  corresponds to a value which is difference between maximum and
minimum of indices.  Figure 4 displays wavelet cross correlations of BSE
and NSE indices. It is evident from this figure that for smaller scales
(i.e. between 1000 and 2700 approximately )the indices  have negative correlation.
Very strong anti-correlation corresponds to value of 2100. This scaling value
of 2100 corresponds to difference between average value of BSE index and
average value of NSE index. 

\begin{figure}
\begin{center}
\includegraphics[angle=270,width=7.0cm]{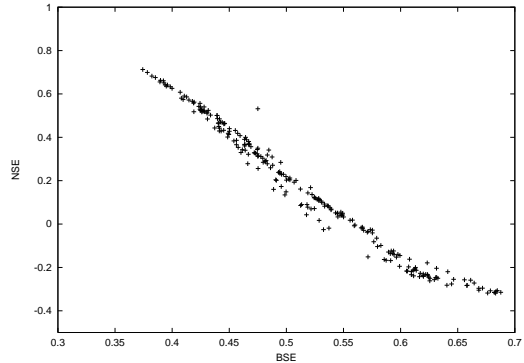}
\caption{Scatter of wavelet coefficient of BSE  and BSE indices at scale = 2000} 
\end{center}
\end{figure}

In figure 5 we have plotted wavelet coefficient w of BSE and NSE indices
for scaling value of 2000.  Figure 6 has been obtained by plotting wavelet coefficient
of BSE and NSE index for scaling value of 6000.
From figure 5 it is clear that data scatter corresponds to highly anti-correlated value
and in figure 6 data scatter depicts strong correlated value.
While in figure 2 data scatter represents classical picture of
correlated scatter but in figures 5 and 6 
wavelet coefficients corresponding to BSE and NSE indices have been plotted 
and represent anti-correlated scatter and correlated scatter respectively.
Figures 5 and 6 confirm the observations made from figure 3 that BSE and NSE indices
can be highly correlated or less correlated or anti-correlated depending on scale.

\begin{figure}
\begin{center}
\includegraphics[angle=270,width=7.0cm]{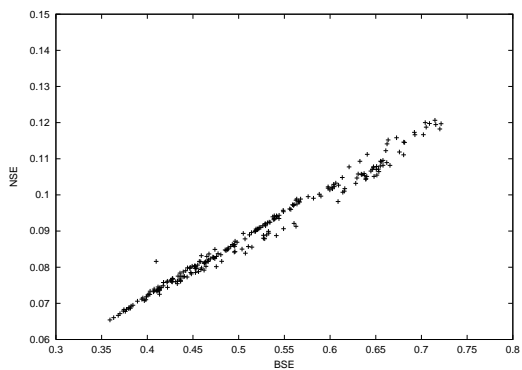}
\caption{Scatter of wavelet coefficient of BSE  and BSE indices at scale= 6000} 
\end{center}
\end{figure}

Studies involving
relationships between various stock exchanges have become important because
of globalization. It is very complex problem to study influences and links
between various stock exchanges because stock exchanges world over open and
close at different times and each individual stock exchange, trades in
portfolios which may be unique to its local economy and nature of currency involved
is also different.
Yet it has been possible to [16] to study taxonomy of
various stock indices. It has been shown that open-closure return of New York
stock exchange at day t and open-closure return of Tokyo stock exchange at
day t+1 are strongly correlated [17]. A detailed study involving quasi-synchronous
correlation coefficient between indices of various markets in the world
have been done to evolve hierarchical tree [16]. Using correlation coefficient
Mantegna et al [6,18] have studied minimum spanning tree of economical data.
From MST it is possible to obtain ultra metric space and hierarchy of various
indices. MST approach has been earlier used to describe various complex systems
like spin glass etc. MST results have shown that portfolios of technology
,software ,energy, food etc form clusters and each cluster turns out
to be branch of MST tree [18] and this clustering sometimes extends even
to global scale. Vandewalla et al [19] extended this studies by investigating
specific topology of financial MST where it has been shown that
markets follow   critical
self organized topology.  This hierarchical structure is believed to result
in cascade of information and clusters of buy-sell
orders [20,21] and sometimes to crashes [22]. It has been shown that tree
length shrinks during a stock market crisis [23].The time dependent properties
of minimum spanning tree gives deep insights  in market dynamics [24]
 because markets have been characterized
as evolving  complex system [25].

Correlation coefficient forms an important input to study taxonomy, MST or time dependent 
dynamics of markets. All such studies have been carried out using equation (1) which 
characterizes data at global level.
But interest lies in studying  correlation
ratio from one scale to another scale which has been achieved by defining
wavelet based correlation coefficient. By using wavelet scale dependent correlation ratio
it may be interesting to study taxonomy, MST etc. which may give further insights in market dynamics.

\noindent{\bf Conclusions:}

By using wavelet concepts in this paper, we have shown that  correlation 
between BSE and NSE indices is scale dependent. They are   
strongly correlated or less correlated or strongly anti-correlated depending
on scale.  
 
\noindent{ \bf References: }
\begin{enumerate}
\item  N.Vandewalle, M. Ausloos in Econophysics -an Emerging Sciences
       Edited by J.Kertesz,I.Kondor (Kluwer,Dordrecht 1999 Edition)
\item  V.M.Yakovenko, cond-mat/0302270,13feb 2003
\item  Ashok Razdan,Pramana -Journal of Physics 58(2002)537
\item  M. Ausloos and K.Ivanova, cond-mat/0108394,24 aug 2001
\item  E.F.Femma, J.Finance (1970) 383-417
\item  R.N.Mantegna, H.E. Stanely , An Introduction to Econophysics
       Cambridge University Press ,2000.
\item  E.E.Peters, Chaos and Order in Capital Markets, Wiley,New York,1991
\item  E.E.Peters,  Fractal Market Analysis,Wiley,New York,1994
\item  S.Ghashghaie, W.Breymann,J.Pineke, P.Talkner and Y.Dodge, Nature 381(1996)767
\item  R.N.Mantegna and H.Stanley, Nature 383(1996)587
\item  N.Vandewalle and M.Ausloos, Physica A 246(1997)454
\item  I.Daubechies,Commun. Pure Appl. Math., 41(1988)909
\item  I.Daubechies , Ten Lectures in Wavelets 1992, Vol 61 CBMS-NSF series in Applied Mathematics
\item  A.Razdan,A.Haungs,H.Rebel and C.L.Bhat, Astroparticle Physics 17(2002)497,
       Astroparticle Physics 12(1999)145
\item  P.Frick, R.Bcek, E.M.Berkhuijsen and I.Patrickeyev, Astro-ph/0109017,3 sept 2001
\item  G.Bonanno, N.Vandewalle,R.N.Mantegna ,cond-mat/0001268 11 aug 2000
\item  K.G.Becker,J.E.Finnerty and M.Gupta, Journal of Finance XLV (1990)1297
\item  R.N.Mantegna, Eur. Phys. J B 11(1999)193
\item  N.Vandewalle , F.Brisbois and X. Tordoir ,cond-mat/0009245 16 sept 2000
\item  A. Arneodo ,J.F.Muzy and D.Sornette, Eur.J Phys. B 2(1998)277
\item  M.Pasquini and M.Serva, Eur. J.Phys. B 16(2000)195
\item  E.Canessa , J. Phys. A: Math. Gen. 33(2000)3637
\item  J.-P.Onnela, A. Chakraborti,K.Kaski and J.Kertesz,cond-mat/0212037,2 Dec 2002
\item  J.-P.Onnela, A. Chakraborti,K.Kaski,J.Kertesz and A.Kanto,cond-mat/0302546,26 Feb2003
\item  W.B.Arthur,S.N.Durlauf and D.A.Lane (eds.), The economy as an evolving
       complex sustem II, Addison Wesley (1997)
\end{enumerate}
\end{document}